\documentclass{article}

\usepackage{arxiv}

\usepackage[utf8]{inputenc} 
\usepackage[T1]{fontenc}    
\usepackage{hyperref}       
\usepackage{url}            
\usepackage{booktabs}       
\usepackage{amsfonts}       
\usepackage{nicefrac}       
\usepackage{microtype}      
\usepackage{lipsum}
\usepackage{graphicx}
\graphicspath{ {./images/} }

\title{Designing learning experiences for online teaching and learning}

\author{
 Nachamma Sockalingam \\
  Learning Science Lab\\
  Singapore University of Technology and Design\\
  nachamma@sutd.edu.sg
  \And
 Junhua Liu \\
  Information Systems Technology and Design\\
  Singapore University of Technology and Design\\
  junhua\_liu@mymail.sutd.edu.sg
}

\begin{document}
\maketitle
\begin{abstract}
Teaching is about constantly innovating strategies, ways and means to engage diverse students in active and meaningful learning. In line with this, SUTD adopts various student-centric teaching and learning teaching methods and approaches. This means that our graduate/undergraduate instructors have to be ready to teach using these student student-centric teaching and learning pedagogies. In this article, I share my experiences of redesigning this teaching course that is typically conducted face-to-face to a synchronous online course and also invite one of the participant in this course to reflect on his experience as a student. 
\end{abstract}


\section{Introduction}
Teaching is about constantly innovating strategies, ways and means to engage diverse students in active and meaningful learning. Each and every lesson is dynamic. Even without the unprecedented situation of COVID-19 and lockdown of educational institutions, we recognize that teaching cannot be just didactic and passive lecturing. While knowledge transfer/ acquisition is necessary, we now understand that our learners need to go beyond and be future-ready so that they are able to solve problems, think critically, work in diverse teams collaboratively, be techno-savvy and be flexible to meet the unforeseen demands of the Volatile, Uncertain, Complex and Ambiguous (VUCA) world.

In line with this, SUTD adopts various student-centric teaching and learning teaching methods and approaches. For instance, SUTD uses team teaching; that is a team of faculty instructors, graduate/undergraduate teaching assistants come together to
teach a cohort of 50 students. This ensures that the faculty to student ratio is kept low (from 1:11 to 1: 16 ratio) and makes it possible for students to get individualized attention.

This means that our graduate/undergraduate instructors have to be ready to teach using these student student-centric teaching and learning pedagogies. Even though there is a common belief that only the gifted can teach, we take the perspective that teaching can be learnt by anyone if we try to gain a deeper understanding of what teaching entails. We recognize the need to support our graduate teaching assistants in preparing to teach and run a strategic 18 hour, 6-week course to prepare them to be graduate teaching assistants so that it benefits them as instructors, the faculty instructors they work with and also their students.

In this article, I share my experiences of redesigning this teaching course that is typically conducted face-to-face to a synchronous online course and also invite one of the participant in this course to reflect on his experience as a student.

\section{Pedagogical frameworks / models underpinning the redesign of online course}

The “Teaching@SUTD” course is typically conducted face-to-face and embraces various active learning teaching methods\cite{b1}. However, we had to conduct this course completely online due to the lockdown situation. This course was run from 18 May 2020 to 22 June 2020 for a batch of 25 students.

\begin{figure}[t]
\centering
  \includegraphics[width=0.75\textwidth]{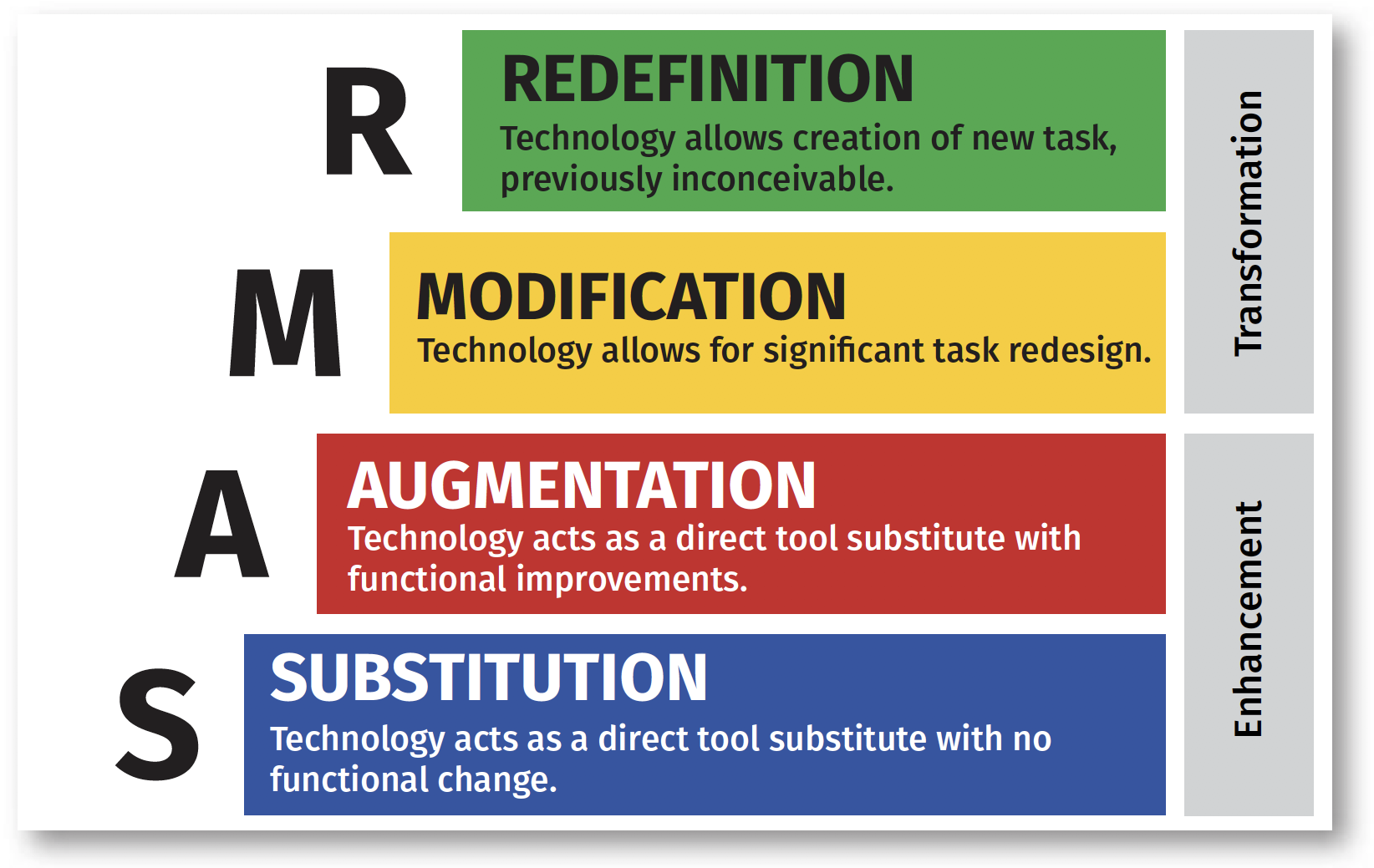}
  \caption{The Substitution, Augmentation, Modification and Redefinition (SAMR) model.}
  \label{fig:f1}
\end{figure}

Given the limited time and resources and quick turnaround time, I decided to “Augment and modify” the course instead of completely “redefining” the course as per Ruben Putendra’s SAMR model which stands for Substitution, Augmentation, Modification and Redefinition (Figure~\ref{fig:f1}). The SAMR model provides a classification of the various types of transformation of face-to-face to blended/online lessons\cite{b2}.

Typical online teaching and learning lessons that we experience such as that in corporate trainings, or even Massive Open Online Courses (MOOC) courses refer to provision of structured reading materials, audio and video resources followed by assessments such as quizzes or assignments for independent and self-directed learning. The advantages of such a mode of online teaching and learning are the conveniences of learning anytime, anyplace, anywhere, being able to revisit the learning materials repeatedly, cheaper cost to students and the wider reach to audience.

However, this mode of online teaching and learning is meant for self-directed learning and mainly knowledge acquisition, and may not cover humanistic aspects such as collaboration and communication skills when used solely in an asynchronous/self-directed learning mode. Hands-on sessions such as lab work or practical work will also be limited. So it important that when we redesign our teaching practices for online learning, we do not simply substitute our mode of delivery to be online, but consider the purposes of our activities and the learning outcomes from each of the activities to “modify” or ideally “redefine” the lessons.

In my case, I needed my students to be able to deliver an online/blended learning that incorporated SUTD’s active and interactive learning in groups. So, I decided on “ Augmentation and Modification” depending on the activities. While the SAMR model is useful for classifying the transformation type, it does not guide us in what factors we need to consider in the redesign process. The “Fit for purpose” teaching and learning design framework for blended/online teaching and learning \cite{b3} I had developed addresses this and helps to plan the technology tools for online teaching.

\begin{figure}[t]
\centering
  \includegraphics[width=0.6\textwidth]{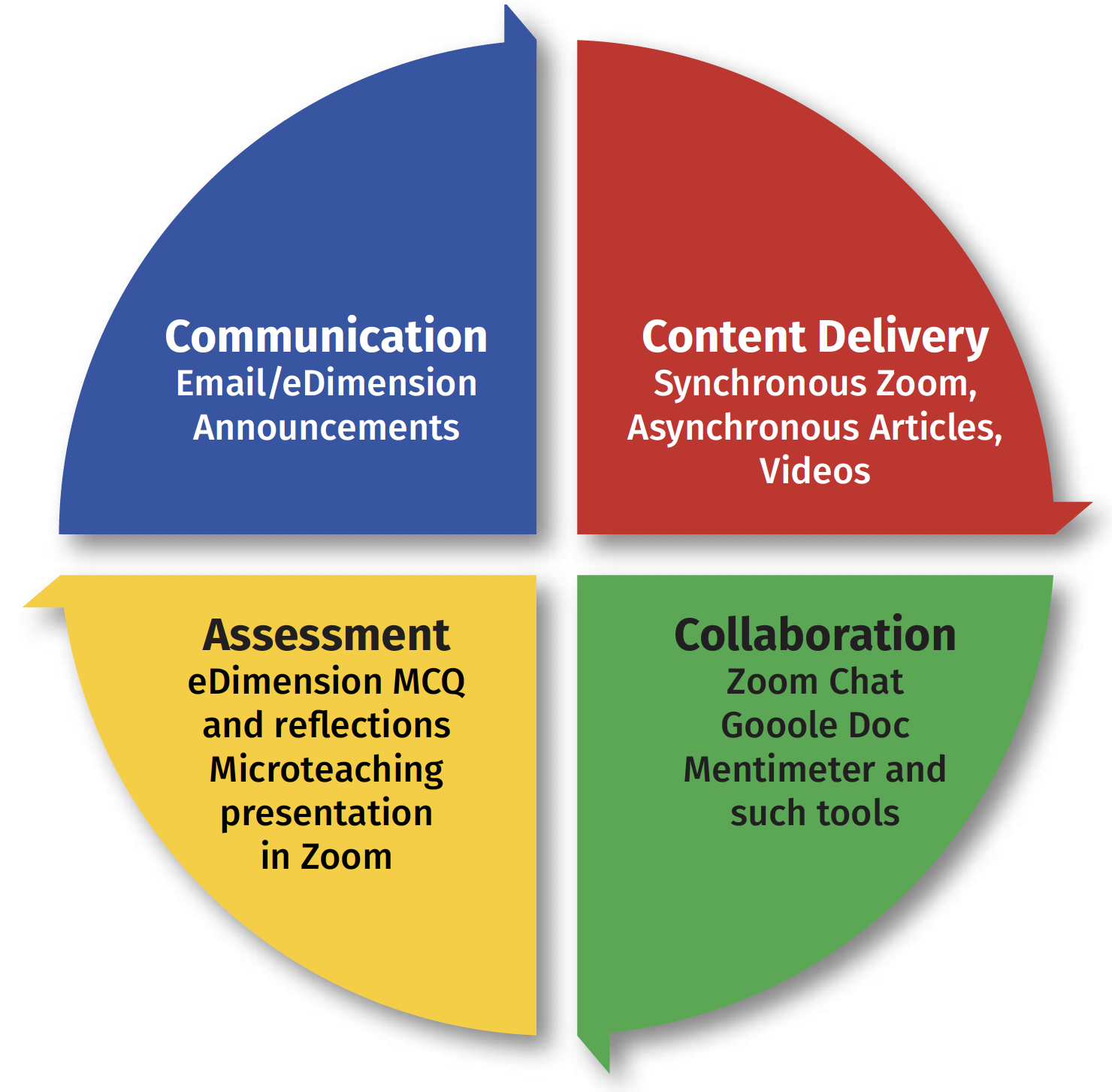}
  \caption{"Fit for purpose" teaching and learning design framework for blended/online teaching}
  \label{fig:f2}
\end{figure}

The “Fit for Purpose” redesigning framework basically breaks the teaching and learning activities into purposeful chunks and proposes that we select technology tools according to the purposes of the teaching and learning activity chunks (Figure~\ref{fig:f2}). The framework considers various factors like (i) Learning outcomes, (ii) Teaching and learning activities, (iii) Assessment, and (iv) Appropriate technology tools in redesigning our lessons, and it can even help us in considering the sequence of our lessons. It combines various concepts such as Bloom’s Taxonomy \cite{b4}, Constructive Alignment \cite{b5}, Backward design\cite{b6}, TPACK model \cite{b7} in one framework. The recommendation is that we use the framework for individual lesson plan and build up to the module plan. This framework is also shared with our teaching faculty in our EPTL website and you will see the examples of implementation in various courses.

\section{Redesigning the Teaching@SUTD course for online learning}

\begin{figure}[t]
\centering
  \includegraphics[width=\textwidth]{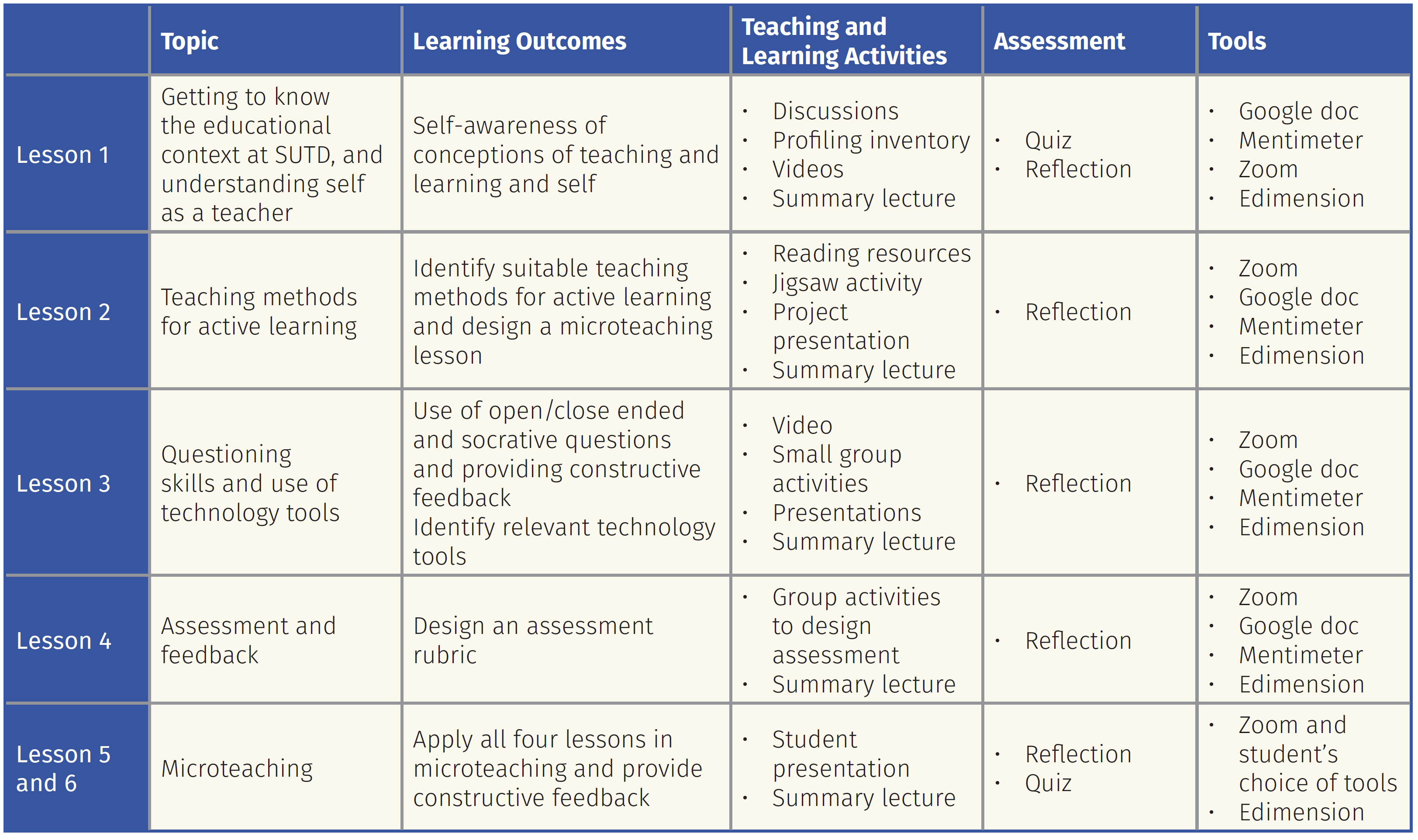}
  \caption{An overview of the online "Teaching at SUTD" course using the Fit for Purpose" Framework}
  \label{fig:f3}
\end{figure}

Let us now see how the framework was used. Figure~\ref{fig:f3} gives a quick overview of the Teaching at SUTD course and maps it to the “Fit for purpose” framework.

SUTD’s Learning Management System, eDimension, and emails were the primary mode for Classroom Management and Communication. Content Delivery was conducted either synchronously using video conferencing tool Zoom or asynchronously using YouTube videos/reading materials in eDimension. Collaboration and Group Work amongst students were done using video conferencing tool Zoom chat, online word processing tool Google Doc as well online Interactive presentation tool Mentimeter in Zoom sessions. Google Doc was useful as it allowed for collaborative and simultaneous editing and writing. Mentimeter allowed for interactive question and answers for quick polls, quizzes and icebreaking activities. Both formative and summative assessments were conducted using tools such as online synchronous quizzes in eDimension, open-ended reflections in eDimension and student presentations in Zoom with peer and instructor feedback.

\begin{figure}[t]
\centering
  \includegraphics[width=\textwidth]{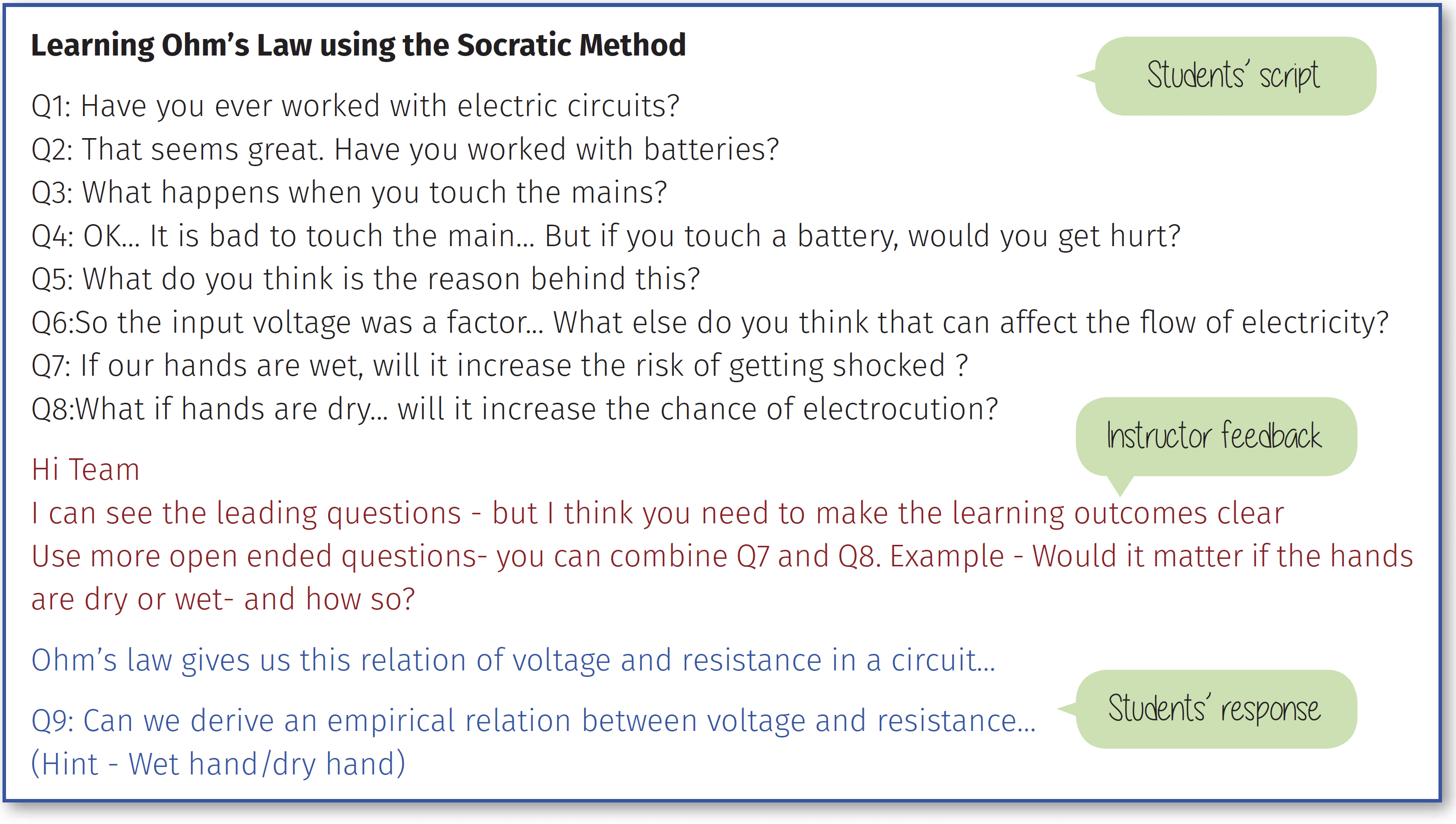}
  \caption{Students'' script to teach using socrative questioning method and instructor feedback with student response}
  \label{fig:f4}
\end{figure}

While we had used tools such as Google Doc and Mentimeter in previous runs of the face-to-face classes as well, we used these tools more extensively in the online course this time, especially in classroom discussion (Figure~\ref{fig:f4}). Typically, this collaborative activity would have been a role play activity of 3 students where one student plays the role of a teacher, another of a student and the third one serves as the observer. The teacher is to teach a certain concept using socrative questioning method, and after the 15 minute activity of planning and executing, students will reflect on their experiences to discuss how the teaching activity could have been improved. They typically use a vanguard sheet for their script. But this time, the team of students wrote out the their socrative questioning in GoogleDoc. I found this useful as I had a concrete script to provide feedback on even after class. It also meant that other student groups could read each other’s scripts and feedback to learn from.

The main difference in terms of the technology use this time was the use of Zoom platform to host the synchronous lessons. We used the breakout rooms in Zoom for small group discussion. However, due to the technical limitation of the breakout room, which only allows the instructors/hosts to move from room to room, we could not simulate one of the activities in Lesson 2 effectively.

This was the Jigsaw activity, where group members rotate and go through 6 stations to peer learn 6 teaching methods to construct and formulate their understanding. To adapt this to online format, and since it was not possible for students to move from one
station to another, we sorted student names into different breakout rooms and asked students to peer teach. In this case, students did not go from station to station; but students from different stations were sorted into one group. While this is technically the same as moving from station to station, we found that this activity was not as effective as in classroom setting due to various reasons. For example, not all students had read up, somewhere not speaking up and some others were a little confused with the activity since they were unfamiliar. Generally, the energy and buzz was also not there.

In my views, this activity had worked better in classroom. This is possibly because there would be an expert stationed in one place, and after repeating the explanation 6 times, they would have identified their gaps and would have been able to fill their gaps upon
returning to their original station and group. In this way, each group would at least knew one teaching method very well and since the others students were learning from experts, they would have picked up the essentials. The advantage was learning in and from groups. But now, everyone had to be an expert and needed everyone’s cooperation. As an instructor, I was also not able to monitor the energy level in the various groups simultaneously and move from group to group to energize them by moving around the classroom space as I would have in a physical classroom.

Interestingly, the other online activities that were mostly non-physical and those that focused on cognitive engagement ( e.g., designing an assessment) went well.

The assessments were kept as the same format as in face-to-face classes although I modified the questions to suit the context this time. For example, the first reflection questioned on what it means to be an online teacher. The synchronicity of lessons, open-endedness in assignment questions, and focus on the thinking process helped to make the assignments suitable for online assessments so that there is originality and copy-pasting is minimized. I did not see any of these.

The final assessment of microteaching was completely online and this was probably the most challenging for the students in my opinion compared to the previous terms. Senior Lecturer Oka Kurniawan from ISTD also facilitated the microteaching sessions as an assessor. Students had to be creative in designing their online teaching activities and teach within 5 minutes. However we did give additional 5 minutes of buffer time since this was online. The students had to focus more on cognitive activities and minimize on physical activities for their microteaching. In the previous years, many students would include physical activities and demos in their microteaching. I recollect a student teacher asking the class members to imagine themselves to be odd/even numbers and group us according to instructions to illustrate concepts on mathematical clustering. This reminded me of Tactile Mathematics. But such activities would have to be redesigned to be taught online and the graduate students rose up to the occasion.

\section{Lessons Learnt}

Overall, while we have the various frameworks and models to redesign the face-to-face to online lessons, this is not a simple process. We have to juggle with the technical and technology limitations and also set new social norms in online teaching and learning. For instance, we had to resort to switching off the online videos to avoid streaming lags but that also meant that we could not keep track of student engagement continuously,
and so we substituted with Zoom chats and emoticons. Also, it took a slightly longer time to build the social connection with the students to gain their trust and understanding, and seek the cooperation. Building the social connection is particularly important in group projects.

To get a better perspective, I sought weekly feedback from students. The learners’ sentiments of the online class varied. While some liked the online activities, some others found the activities a little monotonous and limited. The discussions and collaborative activities were found to be challenging especially in the early weeks when participants were still warming up but became better over the weeks.

Some noted in their reflections that they still prefer face-to-face learning over completely online learning. Even though blended online learning, that is a blend of synchronous and asynchronous online learning may offer some relief over completely self-directed and independent online learning, students seemed to actually prefer a blend of face-to-face and online learning. There were some rare extremes of students
actually preferring completely online (but this misses the importance of learning from peers in collaborative groups). I too agree that the value of personal and social connection in face-to-face meetings is important.

In my personal views, teaching is not all just about just the learning outcomes, skills and values; Many tacit values, habits (e.g., being organized, being positive), mentorship, professional relationships are formed through teacher-student and student-student connections and these are difficult to foster and maintain online.

Overall, the course participants indicated that they felt positive and confident at the end of the course about active teaching and learning in an online environment. All of the graduate students demonstrated active, student centric teaching methods in their
online teaching (even though many were not familiar with such methods at the onset) and some were given additional opportunities to refine their work when needed. We continue to improve the course based on student feedback.

Here are some tips for designing online learning experiences based on my experience.
\begin{enumerate}
  \item Establish communication channels and connect with students for better understanding and relationships.
  \item Set expectations on social norms for online learning and communicate effectively on what is expected. E.g., Explain what a Jigsaw activity is.
  \item Do not just Substitute face-to-face lesson to online mode – Augment, Modify or Redefine (SAMR model) activities in redesigning the lessons as needed. You can redesign gradually instead of one-shot if you have time limitations.
  \item Redesign using “Fit for purpose” activities and tools.
  \item Minimize physical activities if technology is limiting and redesign activities. For instance, consider virtual tools for labs and reflections to focus on process skills.
  \item Blend synchronous and asynchronous activities, and connect the synchronous and asynchronous activities.
  \item Make assessments open-ended and focus on process skills.
  \item Use a good mix of formative and summative assessments, and provide prompt feedback, involving the peers and even industrial experts.
  \item Give opportunities for students to resubmit work where possible.
  \item Get informal feedback from students and continually modify where possible.
\end{enumerate}

This experience of going completely online in such a short time frame was indeed positive and fruitful. I really appreciate and value my students’ engagement and learnt from them as well. Their constructive comments on improving the course gave me refreshing ideas. Reflecting through this article also helped me to decide on which activities to keep and modify for the next run.

I end the article with one of the reflections by our graduate student Liu Jun Hua, who is an AI entrepreneur with deep interest in teaching and learning. We both concur that focusing on the learner and connecting with the learner is the most element for successful teaching - especially in an online context.

\section{Reflection as a student}

The 6-week GTA course was conducted during the COVID-19 outbreak, where physical classes were not possible. It was the first time that I participated in a class completely online. Fortunately, the online classes went well. I had great joy going through the materials and interacting with the instructor and teachers on Zoom. Most of the students participated in the discussions and class activities actively.

Through the microteaching experience, I realised that it is certainly a challenging task to prepare class materials for online classes that help students achieve learning goals while encouraging active participation. From the entire course and the final microteaching sessions, I observed some patterns or methods that the various presenters had used in engaging the audience and encouraging participation. These are summarized as follows.

\textbf{Focus on the audience, not the content}.
Pushing all the prepared content to the audience is perhaps the easiest way to conduct a class, but not necessarily the most effective one. Focusing on the audience is really important, especially for live classes. Be aware of whether majority of the audience are following and being engaged. For instance, are they paying attention? Are they responding to my probing? Are they thinking and taking notes? Learning online can be distracting. Therefore, it is crucial that the instructor keeps in mind students’ attention during the class.

\textbf{Orchestrate the flow of content}.
Designing a class is like planning for a music performance — we need to anticipate the emotion of the audience though the whole play to create an impact\cite{b9}. We can variate the rhythm and intensity of the content to orchestrate the audience’s emotion to keep them engaged.

\textbf{Use of real-world examples}.
While knowledge is developed constructively, connecting abstract and non-trivial content to real-world examples will tremendously help the audience create connection and see the purpose of learning~\cite{b10}. Furthermore, making the content light-weighted and fun will certainly help with the engagement — who doesn’t like humour?

\section{Acknowledgement}

The article is published on the EduSCAPES: AN SUTD PEDAGOGY MAGAZINE, page 49-53.

\bibliographystyle{unsrt}


\end{document}